\documentclass[aps,prb,twocolumn,groupedaddress]{revtex4}
\usepackage{epsfig}
\begin{document}

\title{On the nature of the magnetic ground-state wave function of V$_2$O$_3$. }
\author{S. Di Matteo$^{1,2}$}
\affiliation {$^{1}${Dipartimento di Fisica "E. Amaldi", Universit\`a di Roma 3, via della Vasca Navale 84, I-00146 Roma, Italy}\\
$^{2}${Laboratori Nazionali di Frascati INFN, via E. Fermi 40 I-00044
Frascati, Roma, Italy}}
\date{\today}

\begin{abstract}
After a brief historical introduction, we dwell on two recent experiments in the low-temperature, monoclinic phase of V$_2$O$_3$: K-edge resonant x-ray scattering and non-reciprocal linear dichroism, whose interpretations are in conflict, as they require incompatible magnetic space groups. Such a conflict is critically reviewed, in the light of the present literature, and new experimental tests are suggested, in order to determine unambiguously the magnetic group. 
We then focus on the correlated, non-local nature of the ground-state wave function, that is at the basis of some drawbacks of the LDA+U approach: we singled out the physical mechanism that makes LDA+U unreliable, and indicate the way out for a possible remedy. Finally we explain, by means of a symmetry argument related to the molecular wave function, why the magnetic moment lies in the glide plane, even in the absence of any local symmetry at vanadium sites. 

\end{abstract}

\maketitle

\section{Introduction}

In the last 35 years V$_2$O$_3$ has been the subject 
of wide experimental and theoretical investigations, after its identification as the prototype of Mott-Hubbard systems \cite{rice}.
At the beginning of the seventies an intensive set of measurements (resistivity \cite{mcwhan1}, specific heat \cite{mcwhan2}, magnetic susceptibility \cite{mcwhan3}, x-rays \cite{dernier}, and polarized neutron diffraction \cite{moon}) was performed in order to clarify its crystal and magnetic structure as well as its transport properties.  This marked a break-point with all previous theories \cite{adler1,rice1} that described the "metal-semiconductor" phase transition of V$_2$O$_3$.
Actually, two metal-insulator transitions were found: one from a paramagnetic metallic (PM) to an antiferromagnetic insulating (AFI) phase, below $\simeq 150$ K, associated with a corundum-to-monoclinic crystal distortion; the other from the PM to a paramagnetic insulating (PI) phase at higher temperatures (above $\simeq 500$ K): among transition metal oxides it is the only known example of PM-PI transition \cite{weibao}, with no structural deformations and no magnetic degrees of freedom involved. The PI phase can be also obtained by means of a small ($1\div 2 \%$) Cr-doping: in this case Cr ions act as a negative pressure, thus enlarging the average  cation-cation distance and inducing a Mott-insulator transition. As described in \cite{spalek}, all transitions are due to the interplay between band formation and electron Coulomb correlation.

In 1978 Castellani, Natoli and Ranninger (CNR) attempted, in a series of papers \cite{cnr}, a realistic description of the complex magnetic properties and phase diagram of V$_2$O$_3$. CNR focused on the peculiar magnetic structure observed in the AFI phase, that breaks the high-temperature trigonal symmetry and can not be explained in terms of a single-band Hubbard model. They realized that the introduction of the extra degree of freedom represented by t$_{2g}$-vanadium orbitals was necessary to describe correctly the spin structure. The driving mechanism was the ordered pattern of these t$_{2g}$-orbitals throughout the whole crystal, the so-called "orbital ordering". Twenty years later, the advent of modern third generation synchrotron radiation sources stimulated an interesting prediction \cite{fabrizio} to look for the expected orbital ordering by means of resonant x-ray scattering (RXS). Soon after, its experimental detection was claimed to be found \cite{paolasini}. The subsequent theoretical and experimental debate led to the abandonment of the CNR model, that was based on a spin $S_V=1/2$ on each vanadium ion and thus in contradiction with the experimental result of $S_V=1$ \cite{paolasini,park}, in favour of a picture where the two nearest V-ions are linked together in a stable molecule with spin $S_M=2$ \cite{mila12,dimatteo,tanaka}. This picture has been recently criticized by Elfimov {\it et al.} \cite{elfimov}. Yet, the LDA and LDA+U schemes at the basis of this work are unable to describe the physics of V$_2$O$_3$, due to a fundamental weakness of local density approximations to mimic the behaviour of the correlation potential, as we shall see in Section 4. Thus, I believe that, especially in the version of Tanaka \cite{tanaka} that includes the spin-orbit interaction, the "molecular" model catches the main energy scales and correctly describes many relevant ground state properties: spin magnitude and direction, softening of the C$_{44}$ phonon mode, spin to orbital moment ratio and relative orbital anisotropies ($a_{1g}$ to $e_g$ ratios) in the three PM, PI, and AFI phases. 

The paper is divided as follows: in Section 2 the implications of the RXS experiments \cite{paolasini,paolasini1} about the nature of V$_2$O$_3$ magnetic space group are introduced. Section 3 is devoted to a different interpretation of the "non-reciprocal" linear dichroism experiment in Ref. \cite{goulon}, in order to reconcile it with the magnetic space group as determined by RXS. Finally, in section 4, the breakdown of both local density approximation and LDA+U is illustrated, due to the non-local correlation potential generated by the molecular ground-state wave function. The molecular ground state is also at the basis of a simple explanation, based on symmetry arguments only, of the reason why the magnetic moment should lie in the glide-plane, in spite of the absence of a local symmetry element at vanadium sites.

\section{Interpretation of resonant x-ray scattering experiments.}

The RXS experiments \cite{paolasini,paolasini1} were predicted to occur on the basis of CNR model, and had of course to be reinterpreted once the failure of such a model had been recognized. Such a task was undertaken by Tanaka \cite{tanaka} and Lovesey and collaborators \cite{lovesey}, who, independently, identified the RXS signal as a signature of the magnetic distribution and not of the orbital ordering of vanadium electrons. 
Their interpretations are based on the commonly accepted magnetic space group (MSG), $P2/a+{\hat T}\{E|t_0\}P2/a$, derived from x-ray \cite{dernier} and polarized neutron \cite{moon} experiments, that contains explicitly the time-reversal symmetry ${\hat {T}}$. Here $t_0$ is the body-centered translation and $P2/a$ the monoclinic group containing the identity  ${\hat {E}}$, the inversion ${\hat {I}}$, the two-fold  rotation about the monoclinic b$_m$-axis ${\hat {C}}_{2b}$ and the  reflection ${\hat {m}}_b$ in the plane perpendicular to this axis. Of course, the
appropriate translation is associated to each of these operators, as detailed below:

\begin{eqnarray}
\begin{array}{lll}
1){\rm ~} {\hat E}, {\rm ~} {\hat I} &\rightarrow & {\rm No ~translation} \\
2){\rm ~} {\hat C_{2b}}, {\rm ~} {\hat m_b} &\rightarrow &
\frac{1}{2}(\vec{b}_m+\vec{c}_m) \\
3){\rm ~} {\hat T}, {\rm ~} {\hat T \hat I} &\rightarrow &
\frac{1}{2}(\vec{a}_m+\vec{b}_m+\vec{c}_m)\equiv t_0 \\
4){\rm ~} {\hat T \hat  C_{2b}}, {\rm ~} {\hat T \hat m_b } &
\rightarrow & \frac{1}{2}\vec{a}_m ~.
\end{array}
\nonumber
\end{eqnarray}

\begin{figure}
\vspace{-4cm}
\epsfysize=100mm
\centerline{\epsffile{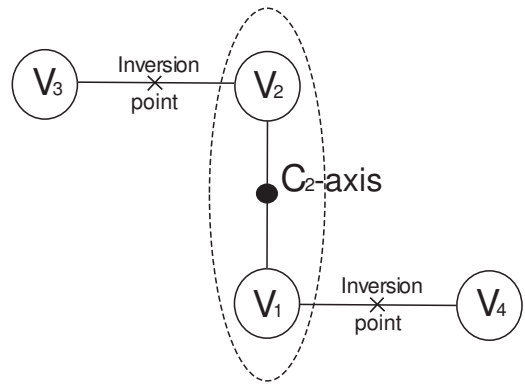}}
\caption{Four vanadium ions lying in the $a_m$-$c_m$ plane are shown. They are all ferromagnetically-coupled and related one another by the four symmetry operations of the $P2/a$-group. The dashed ellipse marks the vertical molecule.}
\label{fig1}
\end{figure}

The origin of the axes has been chosen in the inversion center shown in Fig. \ref{fig1}. Referring to the figure labels, V$_1$ and V$_2$ are connected by the ${\hat {C}}_{2b}$-rotation through their midpoint, V$_2$ and V$_3$ are related by inversion and V$_1$ and $V_3$ by the glide-plane operation ${\hat {m}}_b$, within the plane of the figure. The unit cell contains eight V-ions: the other four are related to these by the time-reversal operation and the body-centered translation. 
Introducing the atomic scattering factor (ASF) as:

\begin{equation}
A \propto \sum_n \frac{\langle \psi_g| {\hat O}^{\dagger} |\psi_n\rangle\langle \psi_n| {\hat O} |\psi_g\rangle}{\hbar \omega- (E_n-E_g) -i\Gamma_n}
\label{asf}
\end{equation}

\noindent the RXS amplitude can be written as: $A\propto\sum_{j=1}^8 e^{i\vec{Q}\cdot\vec{R}_j}f_j$
where the sum is extended over the 8 inequivalent vanadium ions of the magnetic unit cell, at positions $\vec{R}_j$, whose ASF is $f_j$. Here $\psi_g$ and $\psi_n$ are the ground state and the photoexcited state, whose energies are $E_g$ and $E_n$, respectively; the operator ${\hat O}$ represents the matter-radiation interaction, $\hbar \omega$ is the photon energy, $\Gamma_n$ the inverse lifetime of the excited states and $\vec{Q}\equiv(h,k,l)$ is the exchanged Bragg-vector. Reflections with $h+k+l=$odd and $h=$odd \cite{paolasini,paolasini1} are those of interest.
Taking into account of the tensorial character of $f_j$ \cite{dimatteo,lovesey,carra} and the properties of the MSG $P2/a+{\hat T}\{E|t_0\}P2/a$, it is possible to express all the ASF in terms of just one of them, through the appropriate symmetry operations \cite{nota5}. We get:  

\begin{equation}
A=(e^{i\alpha}+\hat{I}e^{-i\alpha})(1-\hat{T}\hat{m}_b)(1-\hat{T})f_1
\label{rxs2}
\end{equation}

\noindent where $\alpha\equiv2\pi(hu+kv+lw)$, and $(u,v,w)$ are the fractional coordinates of vanadium ions in the unit of monoclinic axes. Equation (\ref{rxs2}) has two important consequences: first, it shows that all reflections of kind $h+k+l=$odd and $h=$odd have a magnetic origin, in such a way that the factor $(1-\hat{T})$ be non-zero. In second place, it indicates that they are sensitive to magnetic multipoles of higher order than the magnetic moment, because of the factor $(1-\hat{T}\hat{m}_b)$. This latter observation stems from the fact that the magnetic moment lies on the glide plane $\hat{m}_b$, thus being an eigenvector for $\hat{T}\hat{m}_b$, with eigenvalue +1. Such a picture clearly shows that the presence of the time-reversal symmetry plays a crucial role in the determination of the RXS origin.

Yet, little before the publication of \cite{tanaka} and \cite{lovesey}, Goulon {\it et al.} \cite{goulon} had interpreted the results of their linear dichroism experiment in the monoclinic phase of  V$_2$O$_3$ as an evidence of the non-reciprocal gyrotropic effect, whose implications are in striking contradictions with the previous conclusions.
In fact, a necessary condition to detect the non-reciprocal gyrotropy tensor is that the system be 
magnetoelectric, implying, in turn, that neither time-reversal nor inversion symmetry can separately belong to the MSG.
If the linear dichroism experiment had correctly been interpreted, this would have implied a reduction of $P2/a+{\hat T}\{E|t_0\}P2/a$ to a subgroup not containing ${\hat T }$ and ${\hat I }$, like $({\hat E}, {\hat T \hat I}, {\hat T \hat  C_{2b}}, {\hat m_b})$ or $({\hat E}, {\hat T \hat I}, {\hat  C_{2b}}, {\hat T  \hat m_b})$. 

Thus, one might wonder whether the RXS experiment could be explained with similar arguments as above even in the framework of the above magnetoelectric space groups, in order to be consistent with non-reciprocal effects. My answer is negative, and is motivated by the results of Refs. \cite{dimatteo,dimatteo2,dimatteo3,yvesprb}, that I summarize here.

For practice, we shall distinguish two physical mechanisms that can lead to a MSG reduction. The first is of "charge" origin, i.e., due to any kind of orbital ordering (not necessarily that of CNR); the second concerns the presence of an out-of-plane component of the magnetic moment directed along the $b_m$ axis. The first mechanism has been definitively ruled out by a numerical simulation of the RXS experiment \cite{yvesprb}, based on the finite difference method \cite{yves1}. In fact, in this case, due to the breakdown of the time-reversal symmetry in Eq. (\ref{rxs2}), for $h+k+l=$odd, $h=$odd reflections a signal would appear at $4p$-energies much bigger than the one at the pre K edge, whereas no such a signal is detected experimentally.
The second possibility has been long debated in the literature \cite{moon,word,yethirai}, and yet is still very controversial. The importance of an out-of-plane component of the magnetic moment is related to the fact that it could break the glide-plane symmetry. Three independent polarised neutron scattering experiments \cite{moon,word,yethirai} were not able either to discard or accept this hypotesis. It is interesting to note here that in a crystal field approach, due to the absence of symmetry at vanadium sites, the magnetic moment would be allowed to have any direction. Nonetheless, this picture turns out to be too simple: the symmetry of the molecular wave-function is such as to force the moment in the glide plane (see Section 4).
In any case, an out-of-plane component would again imply the presence of a signal at the (111)$_m$ reflection in the energy region of the conduction $p$-band, contrary to what experimentally detected. In fact, as described in Refs. \cite{dimatteo,tanaka,lovesey}, the extinction rule that forbids this signal comes from the time-reversed glide-plane symmetry, due to the factor $(1-\hat{T}\hat{m}_b)$. This latter is active only when the magnetic moment lies within the mirror plane. For this reason, we can conclude that the absence of any features at $4p$-energies of $h+k+l=$odd, $h=$odd reflections is an indication that the correct MSG of V$_2$O$_3$ is $P2/a+{\hat T}\{E|t_0\}P2/a$. 

Finally, I would like to underline a relatively minor, but still important difference between \cite{tanaka} and \cite{lovesey}. In Tanaka's paper the RXS signal is a consequence of the dipole-quadrupole (E1-E2) interference, thus sensitive to magnetic, parity-odd physical quantities (like a toroidal moment or a magnetic quadrupole - which of them is left unspecified). On the contrary Lovesey and collaborators were able to fit the experimental results of Refs. \cite{paolasini,paolasini1} in the different case of purely quadrupolar reflections (E2-E2), where the extracted information concerns the magnetic octupole of the system. These two views seem conflicting each other, but probably "in medio stat veritas". In fact, a third more recent result on this subject \cite{yvesprb}, based on ab-initio calculations, shows that both channels can contribute to the global signal: by means of a relativistic extension of the multiple scattering theory, implemented in the FDMNES package \cite{yves1}, the authors of Ref. \cite{yvesprb} have been able to find a fair agreement in energy and azimuthal scans with the experimental results \cite{paolasini,paolasini1}. Both E1-E2 and E2-E2 channels are present in the global intensity and their relative contribution depends basically on the energy and on the azimuthal angle at which the reflection is measured. Yet, the single-particle character of these calculations still leaves open the door to future improvements in the description of the pre K edge RXS.

\section{Interpretation of the linear dichroism experiment.}

As sketched above and already discussed in the literature \cite{dimatteo3}, the interpretations of RXS \cite{tanaka,lovesey} and non-reciprocal dichroism \cite{goulon} in their original forms require incompatible MSG. Nonetheless, I believe that there is a way to explain both experiments within a unified theoretical framework, which consists in a different interpretation of the linear dichroism experiment, as suggested in \cite{dimatteo3}. 

This alternative view is based on the possibility that the measured linear dichroism \cite{goulon} is not a gyrotropic signal, but is related to the electric quadrupole, and thus of pure dipole-dipole (E1-E1) origin. The difference between the physical operators associated with these two kinds of dichroism was first derived in Ref. \cite{dimatoli}, where their formal expression was given. In a reference frame where $z$ is the direction of propagation of the x-ray beams, and $x$ and $y$, respectively, those of the associated polarizations, the pure E1-E1 dichroism is described by the electric quadrupole operator $L_x^2-L_y^2$, while the non-reciprocal (E1-E2) signal is determined by the expectation value of $(L_x^2-L_y^2)\Omega_z-c.c.$. Here $\vec{L}$ is the angular momentum operator and $\vec{\Omega}$ is the toroidal moment \cite{radescu,carraprb}. The ab-initio simulations \cite{dimatteo3}, based on either the finite difference method or the multiple scattering theory, clearly demonstrated the presence of a dichroic signal of E1-E1 origin, the size of which is some percent of the absorption signal. This calculated intensity has the same shape and order of magnitude as the one measured by Goulon {\it et al.} \cite{goulon}. Indeed, apart from the ab-initio calculation results, a dichroic signal of E1-E1 origin and of this order of magnitude had to be expected, as the low-temperature distortion of the corundum hexagonal planes along orthogonal directions, where the dichroism is measured, can be up to 4$\%$. These simple considerations explain the findings of Ref. \cite{dimatteo3} also at a qualitative level.

The reason why Goulon {\it et al.} \cite{goulon} have interpreted the dichroic signal as "non-reciprocal" is mainly due to the fact that they found an inverted signal when their experiment was performed with a reversed applied magnetic field. This seems to imply that the measured dichroic signal has a time-reversal odd character (like that of a non-reciprocal tensor), contrary to the case of the electric quadrupole $L_x^2-L_y^2$. In the alternative explanation given in Ref. \cite{dimatteo3}, the change of sign measured in \cite{goulon}, when the magnetic field is reversed, is interpreted as due to the same electric quadrupole, but in a different twin chrystallographic domain. In fact, V$_2$O$_3$ has the tendency to form big monodomains in its monoclinic phase, in an unpredictable way \cite{yethirai}, and there are some specific angles at which the measured intensity in one of the other two monoclinic domains appears exactly reversed. I shall not go into details here, referring to the original paper \cite{dimatteo3} for technical aspects. But I want to underline the following two points. First of all, this interpretation is coherent with all past failures to detect a magnetoelectric effect in V$_2$O$_3$ \cite{astrov,jansen}, and does not require an {\it ad-hoc} mechanism to justify a magnetoelectric space group in the ground state of V$_2$O$_3$. In second place, such an interpretation can be checked by some very simple experiments, that I shall describe below. I indeed believe that this is the only way to really settle the question.

There are basically three ways to get some insight into the problem. The first, and probably best, is to repeat the experiment performed in Ref. \cite{goulon}, but paying particular care to check the monoclinic twin after each transition to the AFI phase, for example by means of an x-ray-scattering equipment: this would give a full geometrical control of the system. The role of the magnetoelectric annealing \cite{goulon,dimatteo3}, in this sense, should also be analyzed. Another possible experimental test to confirm, or reject, the interpretation given in Ref. \cite{dimatteo3}, is to measure the linear dichroism in the AFI phase in the same geometrical conditions as in \cite{goulon}, ie, with the x-ray direction parallel to the corundum axis, but without any magnetoelectric annealing, that would be irrelevant to the purpose: in this way the E1-E1 origin of the signal can be proved or disproved.
Moreover, performing an azimuthal scan of the dichroism around the corundum axis, it is possible to compare directly the signal to the simulations performed in \cite{dimatteo3}. Finally, another way to unravel the contradictions is to study the dependence of twin formations on the applied electric and magnetic fields by means of an x-ray-diffraction equipment, in order to determine if their presence can influence the phase transition from PI to AFI and establish whether the ground-state or some excited state properties are investigated by this procedure.

In the light of all what stated above, and given the implications of the non-reciprocal effect, if confirmed, for the physics of V$_2$O$_3$, its unambigous experimental determination is in my opinion of the utmost importance.

\section{The ground-state in the AFI phase: correlated magnetic moment and non-local exchange.}

In this section I shall focus on two apparently unrelated facts that are nonetheless founded on a common background: {\it i)}, the reasons why LDA (and related LDA+U) are completely unreliable in the description of the ground-state wave function of V$_2$O$_3$; and {\it ii)}, the correct symmetry analysis to explain why the magnetic moments lie in the glide-plane.
The first task is achieved through a critical review of all Refs. \cite{dimatteo,tanaka,elfimov,ezhov,allen}; the second is attained by simply analyzing the local symmetries as determined by the molecular nature of the ground-state wave function, whose point group does not coincide with that at the vanadium site. Both effects are a consequence of the non-local correlations in the wave-function (whose specific form will be precised below). 

Before starting the theoretical analysis of the ground-state wave function, I prefer to introduce the experimental results at the basis of the molecular model, that are probably much more indicative \cite{allen,good}.
In 1976 J.W. Allen \cite{allen} wrote that "The magnetic and optical properties of all the phases of V$_2$O$_3$ show a loss of V$^{3+}$-ion identity". This statement was based on the optical experimental study of (V$_2$O$_3$)$_x$(Al$_2$O$_3$)$_{1-x}$ mixtures ($0.01 \leq x \leq 0.08$). The paper continues with: "Optical studies [...] show that V$^{3+}$-V$^{3+}$ interactions between second and further near neighbors are too weak to destroy the V$^{3+}$-ion identity. This result suggests the possibility that the basic localized-electrons unit in the antiferromagnetic insulating phase may be the nearest neighbor pairs, with the electrons delocalized within the pair". I think that it can be instructive to compare this statement to the one, of opposite tenor, based on LDA(+U) calculations, from Elfimov {\it et al.} \cite{elfimov}: "[...] we show that there are other hopping integrals which are equally important for the band shape as the integral for hopping between the partners of the pair". Why do LDA (+U) results differ so much from optical data and from all theoretical descriptions \cite{mila12,dimatteo,tanaka} ?
The answer to this question is already contained, implicitly, in the discussion of section V in Ref. \cite{dimatteo}, to which we refer for a quantitative treatment: here we just try to delineate qualitatively the physics behind it, trying to add new elements for a better comprehension. As a result, we shall be able to define clearly the reasons behind the failure of local density approximations to describe the physics of V$_2$O$_3$.

As illustrated in Ref. \cite{dimatteo}, a good starting point to evaluate the energetics of the vertical molecule is through an orbitally-degenerate Hubbard Hamiltonian. We shall discuss {\it a posteriori} why it is possible to consider simply the vertical pair, that is the contested point in Ref. \cite{elfimov}, and why our results do not depend on the approximations that are impicit in the Hubbard model.
The parameters that play the most relevant role are: the Coulomb on-site repulsions $U_1$ and $U_2$, between electrons on the same orbital and on different orbitals, respectively; the Hund's coupling constant $J_H$ and the trigonal field splitting $\Delta_t$. The relation $U_1=U_2+2J_H$ holds with very good approximation. 
In V$_2$O$_3$ every V$^{3+}$-ion is octahedrally surrounded by a cage of oxygens and, thus, the two valence electrons are forced to a three-fold $t_{2g}$ configuration.
The effect of the trigonal field splitting is to lift the three-fold $t_{2g}$-degeneracy and to push higher in energy the $a_{1g}$ state, leaving as a ground state the degenerate doublet $e_{u,v}$. The explicit expression of $a_{1g}$, $e_{u,v}$ in terms of 3$d$-orbital wave-functions can be found, for example, in \cite{tanaka}. The main contribution to the kinetic energy in the molecule comes from the terms describing $a_{1g}$-$a_{1g}$ hopping, $\rho \simeq -0.82$ eV, and $e_{u,v}$-$e_{u,v}$ hopping, $\mu\simeq 0.2$ eV \cite{mattheiss} (notice that they have opposite signs). 
In the high trigonal-field-splitting limit the ground state is given, thus, by $|\Psi_1 \rangle \equiv |e_u e_v;e_ue_v \rangle$ (the two groups of $e_{u,v}$ labels refer to the electron occupancy of one or the other ion in the molecule, e.g., in Fig. \ref{fig1}, $V_1$ or $V_2$). Even for the trigonal splitting $\Delta_t\simeq 0.4$ eV, deduced from Ref. \cite{ezhov}, the ground-state wave function is characterized by a high percentage of $e_g$-doublet composition. Notice that similar results were already obtained in 1978 by CNR, in their second paper \cite{cnr}, through an unrestricted Hartree-Fock + Hubbard U calculation, which is very akin to the modern LDA+U.
A complete description of the molecular behaviour within a big range of variation of parameters can be found in Ref. \cite{dimatteo}. Here we focus on the values given in Ref. \cite{ezhov}, even if $J_H$ is clearly overestimated \cite{dimatteo,jcond} in the light of the atomic calculations of Tanabe and Sugano ($J_H = 0.79$ eV) \cite{tanabe}: $\Delta_t \simeq 0.4$ eV; $J_H\simeq 0.93$ eV; $U_2 \simeq 2.8$ eV.
Consider now the FM {\it correlated} molecular wave function \cite{nota2}, first derived by Mila {\it et al.} \cite{mila12}: 

\begin{equation}
|\Psi_0 \rangle = \frac{1}{\sqrt{2}} (|a_{1g}e_u; e_v e_u \rangle + |e_v e_u;a_{1g}e_u\rangle)\otimes |S_M=2 \rangle
\label{psi}
\end{equation}

Its energy, in our description, is given by: $E_0=\Delta_t-(\rho-\mu)^2/(U_2-J_H)\simeq -0.16$ eV. Notice that the reference energy $E_1=0$ has been assigned to the ground state of the LDA+U procedure $\Psi_1= |e_ue_v;e_u e_v \rangle$: here the absence of $a_{1g}$ electrons gives no trigonal-field contribution and Pauli principle prevents any hopping in the ferromagnetic vertical molecule. One might wonder why is $E_0$ so lower, having used the same parameters as obtained by the local density calculations.
To this aim, consider the wave function $|\Psi_2 \rangle \equiv |e_u e_v;a_{1g}e_u\rangle$.
The associated energy is given by: $E_2=\Delta_t-(\rho^2+\mu^2)/(U_2-J_H)\simeq +0.02$ eV, higher than both $E_0$ and $E_1$. There are $180$ meV of difference \cite{nota1} between the "mean-field" state $|\Psi_2 \rangle$ and the "correlated" state $|\Psi_0 \rangle$, coming from the hopping interference term $2\mu\rho/(U_2-J_H)$. The origin of this energy is due to the correlated structure of $|\Psi_0 \rangle$ that makes possible for an $e_v$-electron on V$_1$ to hop on V$_2$ and simultaneously for an $a_{1g}$ electron on V$_2$ to hop on V$_1$ (and vice-versa), without changing state. This form of superexchange interaction is basically a non-local electronic correlation and, as such, it cannot be taken into account in the framework of local density calculations.
These simple considerations show why LDA+U results fail to catch the ground-state wave function: as non-local correlations are neglected, the interference energy $-2\mu\rho/(U_2-J_H)$ is completely missed ($\simeq$180/2=90 meV per ion !) and the purely $e_{u,v}$ state $|\Psi_1 \rangle$ is found as ground state. 

In this picture, if we even consider that there are three in-plane second nearest neighbors and introduce further nearest-neighbor hoppings, as done in Ref. \cite{elfimov}, still the non-local correlation energy of the vertical molecule represents a relevant (if not the main) contribution. Notice that the corundum geometry is such that further-nearest-neighbors hopping integrals cannot give rise to any other correlated hopping of relevant strenght: for this reason all inter-molecular interactions can be treated in mean field (or local density) approximation. The energy gain due to the correlated molecule is so big that it can be hardly conceived as an artefact of the Hubbard Hamiltonian, or that the corresponding state dissolves when further nearest neighbor hopping-integrals are taken into account, as Ref. \cite{elfimov} seems to suggest, provided, of course, all local and non-local forms of energy are correctly considered. Moreover, it is stable against rather large variations of $\Delta_t$, $U_2$ and $J_H$ around the chosen values \cite{dimatteo}.
When this correlation energy is simply neglected, as in local density approximations, the ground-state wave function is mainly determined by the trigonal splitting $\Delta_t$, that pushes higher the one-particle energy level of the $a_{1g}$-electrons, thus advantaging $e_{u,v}$. In my opinion, this is a signature of the failure of LDA+U to describe the ground state of V$_2$O$_3$. Indeed, the results of LDA+U calculations \cite{ezhov} are lacking for many aspects: the spin exchanges along the different bonds are strongly underestimated when compared to the experimental results \cite{word}, as recognized by the authors themselves. The orbital occupancy of mainly $e_ue_v$ kind does not reproduce the findings of Park {\it et al.} \cite{park} interpreted with a substantial $a_{1g}$-occupancy about 20-25$\%$. Notice that this is in keeping with the experimental evidence from Rubinstein \cite{rubin} that found no detectable anisotropy in the Knight shift of the NMR spectrum, no observable quadrupole splitting in the NMR spectrum, and zero anysotropy in single-crystal susceptibility measurements for various cristalline orientations with respect to the magnetic field. All these findings apparently point to a negligeable trigonal field splitting, ie, an almost cubic crystal field. Only the molecular picture, through the interference effect, is able to reconcile these latter experiments with the sizeable value of the trigonal distortion, because of the average orbital occupancy $a_{1g}=25\%$, $e_{u,v}=37.5 \%$, close to the statistical ratio ($33\%$) of the cubic field.
In the light of the previous discussion, the main statement of Elfimov {\it et al.} \cite{elfimov} looses its strenght: these authors performed a tight-binding fitting of their LDA and LDA+U band structure, and concluded that a proper description requires second, third and forth neighbor hoppings. Yet, the evident drawback of LDA+U correlated description makes such a conclusion very weak. Moreover, they focused on the $a_{1g}$ band only, whereas the three bands should strongly interfere.

My conclusion in this respect is that V$_2$O$_3$ is a very peculiar strongly correlated electron system, where ordinary ab-initio codes may loose their original strenght \cite{elfimov,anisimov,held}. The case of LDA+U is emblematical: this approach was initially conceived to describe the insulating magnetic materials whose energy gap was not determined by the Stoner parameter, but by the Coulomb repulsion $U$ \cite{anisimov}. Even though it works correctly in these cases \cite{anisimov}, it is nonetheless based on a {\it local} exchange and correlation potential, and, as such, it cannot treat properly a system like V$_2$O$_3$, where non-local correlations amount to 90 meV per ion. The same arguments apply to the recent LDA+DMFT \cite{held}, as clear from the following statement, from Keller {\it et al.} \cite{held1}: "[...] this small trigonal splitting strongly determines the orbital ground state of the V-ion obtained from LDA+DMFT calculations [...]". This method, in spite of the progress coming from the DMFT, can handle just the local dynamics of the system, while non-local correlations are still neglected. This is the reason why the orbital nature of the ground state seems to be determined by the trigonal field splitting. 

It may be true that the physical idea of the "weakly" interacting gas of vertical molecules \cite{dimatteo} could be oversimplified and that a different picture for the band structure may be required, as in the original idea of Elfimov {\it et al.} \cite{elfimov}: yet, to this aim, an extension of the correlation potential V$_c$ in a non-local direction is, in my opinion, essential to describe the physics of V$_2$O$_3$. All past experimental results, like Allen's, must be explained by any theory, and not simply dismissed, as seems to be the case of LDA-based approaches.
In order to get reliable results, it is necessary to go back to the original form of Kohn-Sham equations \cite{kohn} and find out a good approximation for the correlation potential $V_c$: some hints in this sense could be taken from the form of the non-local molecular wave function (\ref{psi}). In this way new informations about V$_2$O$_3$ band structure can be extracted, that do not neglect the "correlated hopping" energy, and, probably, can also explain, among the others, Allen's results \cite{allen}. 

All the previous analysis was based on the non-local nature of superexchange correlations, only, and did not take into account the spin-orbit interaction, that, nonetheless, was fundamental to explain the results of RXS \cite{tanaka,yvesprb}. Notice that, even if very important to describe the RXS experiments and to remove the orbital degeneracy, this interaction is about one order of magnitude lower than the correlated hopping ($\simeq 25$ meV) and, thus, it can be introduced adiabatically in the ground-state manifold of $|\Psi_0 \rangle$. In this way, its inclusion, though it removes the orbital degeneracy of $|\Psi_0 \rangle$ \cite{tanaka}, leaves unaltered all the previous conclusions. The ground state is modified in:

\begin{equation}
|\Psi_0' \rangle= \frac{1}{\sqrt{2}} (|a_{1g}e^+;e^+e^- \rangle + |e^+e^-;a_{1g}e^+\rangle) \otimes |S_{Mz}=\pm 2 \rangle
\label{psi4}
\end{equation}

\noindent where $e^+ \equiv -\frac{1}{\sqrt{2}}(e_u+i e_v) $ and $e^- \equiv \frac{1}{\sqrt{2}}(e_u-i e_v)$.
Here the spin part of the ground state has been written explicitly. Such a state is characterized by a global spin $S_M=2$, with $S_{Mz}=\pm 2$. 
As discussed previously, a long debate arose in the literature about a possible component of the spin of V$_2$O$_3$ out of the glide plane, especially for its consequences on the magnetic space group. From a theoretical point of view, on the basis of usual crystal field models, such an out-of-plane component could even be expected: in fact, vanadium ions, in their AFI phase, are characterized by the absence of any point symmetry, thus allowing the local magnetic moment to follow, in principle, any direction. 
Yet, the molecular nature of the ground state plays a fundamental role also in the description of the correct symmetry. In fact, as any physical variable is determined through its expectation value on the ground-state wave function, and as the electronic wave function is "molecular", then the magnetic moment must be invariant with respect to all the symmetry operations of the molecular point group and not of the ion point group.
In other words, we identify the molecular spin $S_M=S_{V_1}+S_{V2}=2$ as the physical variable detected by neutron measurements, even if it is customary to speak about the magnetic moment per ion, $S_V=1$, in an average sense. This interpretation parallels and extends to the four-electron case of the molecule the S$_V$=1 atomic picture: here, of the whole space spanned by the two $S=1/2$-electrons on the vanadium ion, the S$_V$=0 subspace is projected out, and only the spin-1 is considered for the ground-state properties. Similarly, in the molecule, only the $S_M=2$ subspace has a physical reality, and all the other spin-states ($S_M=0,1$) are projected out. 

Based on these premises, we are now ready to repeat the previous symmetry considerations from a different viewpoint. In fact, the electronic ground-state of the molecule now has a local symmetry element, contrary to the case of the single ion: this symmetry element is the ${\hat {C}}_{2b}$-axis shown in  Fig. \ref{fig1}, orthogonal to the molecular axis, and running through the midpoint of the molecule. It follows that the molecular magnetic moment must lie either along the ${\hat {C}}_2$-axis, or perpendicular to it (this latter case arises when the time-reversal operation is associated to the ${\hat {C}}_2$-axis). No "mixed" configurations are possible. Neutron experiments allow us to choose: the magnetic moment lies orthogonal to the ${\hat {C}}_{2b}$-axis and, thus, the correct local symmetry for the molecule is ${\hat T}{\hat {C}}_2$. The same conclusion is reached by considering Tanaka's wave function (\ref{psi4}). In fact, under ${\hat {C}}_{2}$ symmetry, $a_{1g}\rightarrow a_{1g}$, $e^+ \rightarrow e^+$ and $e^- \rightarrow -e^-$ (see \cite{tanaka}, pag. 3), and time-reversal symmetry changes sign to the imaginary unit. Thus, $\langle \Psi_0'| L_M |\Psi_0' \rangle = \langle \Psi_0'|({\hat T}{\hat {C}}_2)^{\dagger} L_M {\hat T}{\hat {C}}_2|\Psi_0' \rangle$ and the molecular angular moment is invariant under ${\hat T}{\hat {C}}_2$ symmetry, so that it must lie on the glide plane and force the spin moment in the same direction.
Notice, in this respect, that when Tanaka speaks of a possible out-of-plane magnetic component he refers to the single-ion moment: when combined throughout the unit cell, all the ions belonging to the same vertical molecule have opposite out-of-plane moments and thus the global molecular spin is always in the glide-plane, as the symmetry constraint requires.

In conclusion, I have shown that the non-local, molecular, wave function can explain different physical phenomena, apparently unrelated and inexplicable in the framework of an average, ionic picture.

I would like to thank Stephen W. Lovesey who strongly supported and stimulated the present work, and Calogero R. Natoli who shared with me his knowledge on V$_2$O$_3$.


\begin{thebibliography}{99}
\bibitem{rice}
D.B. McWhan, T.M. Rice, and J.P. Remeika, Phys. Rev. Lett. {\bf 23}, 1384 (1969)
\bibitem{mcwhan1}
D.B. Mc Whan and J.P. Remeika, Phys. Rev. B {\bf 2} 3734 (1970); see also  
D.B. McWhan, J.P. Remeika, S.D. Bader, B.B. Triplett, and N.E. Philips, Phys. Rev. B {\bf 7}, 3079 (1973)
\bibitem{mcwhan2}
K. Andres, Phys. Rev. B {\bf 2} 3768 (1970)
\bibitem{mcwhan3}
A. Menth and J.P. Remeika, Phys. Rev. B {\bf 2} 3756 (1970)
\bibitem{dernier}
P.D. Dernier and M. Marezio, Phys. Rev. {\bf 2}, 3771 (1970)
\bibitem{moon}
R.B. Moon, Phys. Rev. Lett. {\bf 25}, 527 (1970)
\bibitem{adler1}
D. Adler and H. Brooks, Phys. Rev. {\bf 155}, 826 (1967); J. Feinlieb, and W. Paul, Phys. Rev. {\bf 155}, 841 (1967); D. Adler, J. Feinlieb, H. Brooks, and W. Paul, Phys. Rev. {\bf 155}, 851 (1967)
\bibitem{rice1}
D.B. McWhan, and T.M. Rice, Phys. Rev. Lett. {\bf 22}, 887 (1969)
\bibitem{weibao}
Wei Bao, C. Broholm, G. Aeppli, S.A. Carter, P. Dai, T.F. Rosenbaum, J.M. Honig, P. Metcalf, S.F. Trevino,  Phys. Rev. B {\bf 58}, 12727 (1998);
\bibitem{spalek}
J. Spa{\l}ek, J. Sol. State Chem. {\bf 88}, 70 (1990)
\bibitem{cnr}
C. Castellani, C.R. Natoli and J. Ranninger, Phys. Rev. B {\bf 18}, 4945 (1978); C. Castellani, C.R. Natoli and J. Ranninger, Phys. Rev. B {\bf 18}, 4967 (1978); C. Castellani, C.R. Natoli and J. Ranninger, Phys. Rev. B {\bf 18}, 5001
\bibitem{fabrizio}
M. Fabrizio, M. Altarelli, M. Benfatto, Phys. Rev. Lett. {\bf 80}, 3400 (1998)
\bibitem{paolasini}
L. Paolasini, C. Vettier, F. de Bergevin, F. Yakhou, D. Mannix, A. Stunault, W. Neubeck, M. Altarelli, M. Fabrizio, P.A. Metcalf and J.M. Honig, Phys. Rev. Lett. {\bf 82}, 4719 (1999)
\bibitem{park}
J.-H. Park, L.H. Tjeng, A. Tanaka, J.W. Allen, C.T. Chen, P. Metcalf, 
J.M. Honig, F.M.F. de Groot, G.A. Sawatzky, {\it Phys. Rev.} B {\bf 61}, 11506 (2000)
\bibitem{mila12}
F. Mila, R. Shiina, F.-C. Zhang, A. Joshi, M. Ma, V. Anisimov, T.M. Rice, 
Phys. Rev. Lett. {\bf 85}, 1714 (2000); R. Shiina, F. Mila, F.-C. Zhang, T.M. Rice,  Phys. Rev. B {\bf 63}, 144422 (2001)
\bibitem{dimatteo}
S. Di Matteo, N.B. Perkins, C.R. Natoli,  Phys. Rev. B {\bf 65}, 054413 (2002)
\bibitem{tanaka}
A. Tanaka, Jour. Phys. Soc. Jpn. {\bf 71}, 1091 (2002)
\bibitem{elfimov}
I. Elfimov, T. Saha-Dasgupta, and M.A. Korotin, Phys. Rev. B {\bf 68}, 113105 (2003)
\bibitem{paolasini1}
L. Paolasini, S. Di Matteo, C. Vettier, F. de Bergevin, A. Sollier, 
W. Neubeck, F. Yakhou, P.A. Metcalf, J.M. Honig, Jour. of Electr. Spectr. 
\& Related Phenomena, {\bf 120/1-3}, 1 (2001)
\bibitem{goulon}
J. Goulon, A. Rogalev, C. Goulon-Ginet, G. Benayoun, L. Paolasini, 
C. Brouder, C. Malgrange, P.A. Metcalf, Phys. Rev. Lett. {\bf 85}, 
4385 (2000)
\bibitem{lovesey}
S.W. Lovesey and K.S. Knight, J. Phys. Cond. Matter {\bf 12}, L367 (2000);
S.W. Lovesey, K.S. Knight and D.S. Sivia,  Phys. Rev. B, {\bf 65}, 
224402 (2002)
\bibitem{carra}
P. Carra, and T. Thole, Rev. Mod. Phys. {\bf 66}, 1509 (1994)
\bibitem{nota5}
The ASF (\ref{asf}) is a scalar, and it can be written as a scalar product of two tensors \cite{carra,dimatteo2}. When we state, with a slight abuse of notation, that a symmetry operator acts on the ASF, we mean that just one of the two tensors is transformed accordingly.
\bibitem{dimatteo2}
S. Di Matteo, Physica B {\bf 318}, 321 (2002)
\bibitem{dimatteo3}
S. Di Matteo, Y. Joly, C.R. Natoli, Phys. Rev. B {\bf 67}, 195105 (2003)
\bibitem{yvesprb}
Y. Joly, S. Di Matteo, and C.R. Natoli, accepted for publication in Phys. Rev. B
\bibitem{yves1}
Y. Joly, Phys. Rev. B {\bf 63}, 125120 (2001)
\bibitem{word}
W.B. Yelon, S.A. Werner, R.E. Word, J.M. Honig, and S. Shivashankar, J. Appl. Phys. {\bf 52}, 2237 (1981)
\bibitem{yethirai}
M. Yethiraj, S.A. Werner, W.B. Yelon, and J.M. Honig, Phys. Rev B {\bf 36}, 8675 (1987)
\bibitem{dimatoli}
S. Di Matteo, C.R. Natoli, J.Synchr. Rad., {\bf 9}, 9 (2002)
\bibitem{radescu}
E.E. Radescu, and D.H. Vlad, Phys. Rev. E {\bf 57}, 6030 (1998)
\bibitem{carraprb}
P. Carra, A. Jerez, and I. Marri, Phys. Rev. B {\bf 67}, 045111 (2003)
\bibitem{astrov}
D.N. Astrov, Sov. Phys. JETP, {\bf 11}, 708 (1960)
\bibitem{jansen}
S. Di Matteo, A.G.M. Jansen, Phys. Rev. B {\bf 66}, 100402 (2002)
\bibitem{ezhov}
S.Y. Ezhov, V.I. Anisimov, D.I. Khomskii, and G.A. Sawatzky, Phys. Rev. Lett. {\bf 83}, 4136 (1999)
\bibitem{allen}
J.W. Allen, Phys. Rev. Lett. {\bf 21}, 1249 (1976)
\bibitem{good}
J.B. Goodenough, in {\it Progress in Solid State Chemistry}, edited by H. Reiss (Pergamon, Oxford, England, 1971), Chap. 4, pag. 291
\bibitem{mattheiss}
L.F. Mattheiss, J. Phys.: Condens. Matter {\bf 6}, 6477 (1994)
\bibitem{tanabe}
Y. Tanabe and S. Sugano, J. Phys. Soc. Jpn. {\bf 9}, 766 (1954)
\bibitem{jcond}
S. Di Matteo, N.B. Perkins, C.R. Natoli, J. Phys: Condens. Matter {\bf 14}, L37 (2002)
\bibitem{nota2}
The replacement $e_g^+ \rightarrow e_g^-$ leads to another, degenerate, ground state, that we neglect now as it is irrelevant to our purpose. This kind of degeneracy is lifted by the spin-orbit interaction, as detailed in Ref. \cite{tanaka}. 
\bibitem{nota1}
Of course all numbers are important only as an order of magnitude: the "true" ground-state of the Hubbard Hamiltonian has extra contributions as clear from Ref. \cite{dimatteo}. Here we try to pick up the main points, for clarity, even if to the prejudice of precision.
\bibitem{kohn}
W. Kohn, L.J. Sham, Phys. Rev. {\bf 140A} 1133 (1965)
\bibitem{rubin}
M. Rubinstein, Phys. Rev. B {\bf 2}, 4731 (1970)
\bibitem{anisimov}
V.I. Anisimov, J. Zaanen and O.K. Andersen, Phys. Rev. B {\bf 44}, 943 (1991)
\bibitem{held}
K. Held, I.A. Nekrasov, G. Keller, V. Eyert, N. Bl\"umer, A.K. McMahan, R.T. Scalettar, T. Pruschke, V.I. Anisimov, and D. Vollhardt, Proceedings of the Winter School on "Quantum simulations of complex many-body systems: from theory to algorithms", (2002), Rolduc/Kerkrade (NL), and cond-mat/0112079
\bibitem{held1}
G. Keller, K. Held, V. Eyert, D. Vollhardt, and V.I. Anisimov, cond-mat/0402133, pag. 3


\end{thebibliography}
\end{document}